\documentclass[sigconf]{acmart}

\usepackage{booktabs} 
\usepackage{paralist}
\usepackage[ruled, linesnumbered]{algorithm2e}
\usepackage{amsmath}
\usepackage{url}
\usepackage{multirow}
\usepackage{booktabs}
\usepackage{graphicx}
\usepackage{subfigure} 

\usepackage{balance}

\def\BibTeX{{\rm B\kern-.05em{\sc i\kern-.025em b}\kern-.08emT\kern-.1667em\lower.7ex\hbox{E}\kern-.125emX}}
    
\copyrightyear{2019} 
\acmYear{2019} 
\setcopyright{acmcopyright}
\acmConference[KDD '19]{The 25th ACM SIGKDD Conference on Knowledge Discovery and Data Mining}{August 4--8, 2019}{Anchorage, AK, USA}
\acmBooktitle{The 25th ACM SIGKDD Conference on Knowledge Discovery and Data Mining (KDD '19), August 4--8, 2019, Anchorage, AK, USA}
\acmPrice{15.00}
\acmDOI{10.1145/3292500.3330659}
\acmISBN{978-1-4503-6201-6/19/08}

\begin{document}

\title{Personalized Attraction Enhanced Sponsored Search with Multi-task Learning}


\author{Wei Zhao}
\affiliation{
    \institution{State Key Lab of ISN, Xidian Univ.}
}
\email{ywzhao@mail.xidian.edu.cn}

\author{Boxuan Zhang}
\affiliation{
    \institution{Alibaba Group}
}
\email{boxuan.zbx@gmail.com}

\author{Beidou Wang}
\affiliation{
    \institution{Simon Fraser University}
}
\email{beidouw@sfu.ca}

\author{Ziyu Guan}\authornote{Corresponding author}
\affiliation{
    \institution{State Key Lab of ISN, Xidian Univ.}
}
\email{zyguan@xidian.edu.cn}

\author{Wanxian Guan}
\affiliation{
    \institution{Alibaba Group}
}
\email{wanxian.gwx@alibaba-inc.com}

\author{Guang Qiu}
\affiliation{
    \institution{Alibaba Group}
}
\email{guang.qiug@alibaba-inc.com}

\author{Wei Ning}
\affiliation{
    \institution{Alibaba Group}
}
\email{wei.ningw@alibaba-inc.com}

\author{Jiming Chen}
\affiliation{
    \institution{Zhejiang University}
}
\email{cjm@zju.edu.cn}

\author{Hongmin Liu}
\affiliation{
    \institution{Henan Polytechnic University}
}
\email{hongminliu@hpu.edu.cn}

%
\renewcommand{\shortauthors}{Zhao and Zhang, et al.}

%
\begin{abstract}
We study a novel problem of sponsored search (SS) for E-Commerce platforms: how we can attract query users to click product advertisements (ads) by presenting them features of products that attract them. This not only benefits merchants and the platform, but also improves user experience. The problem is challenging due to the following reasons: (1) We need to carefully manipulate the ad content without affecting user search experience. (2) It is difficult to obtain users' explicit feedback of their preference in product features. (3) Nowadays, a great portion of the search traffic in E-Commerce platforms is from their mobile apps (e.g., nearly 90\% in Taobao). The situation would get worse in the mobile setting due to limited space. We are focused on the mobile setting and propose to manipulate ad titles by adding a few selling point keywords (SPs) to attract query users. We model it as a personalized attractive SP prediction problem and carry out both large-scale offline evaluation and online A/B tests in Taobao. The contributions include: (1) We explore various exhibition schemes of SPs. (2) We propose a surrogate of user explicit feedback for SP preference. (3) We also explore multi-task learning and various additional features to boost the performance. A variant of our best model has already been deployed in Taobao, leading to a 2\% increase in revenue per thousand impressions and an opt-out rate of merchants less than 4\%.
\end{abstract}

%
%


%
\keywords{Sponsored Search; E-Commerce; Multi-task Learning; Personalization}

%

%
\maketitle

\section{Introduction}\label{sec:intro}
Online shopping has become a daily routine for many people. In recent years, E-commerce platforms such as Amazon\footnote{www.amazon.com} and Taobao\footnote{www.taobao.com} have undergone a rapid and significant growth. As reported by the Alibaba Group \cite{ali2018report}, in the fiscal year of 2018 the annual active consumers on its retail marketplaces (including Taobao) reached 552 million. Moreover, more and more people use mobile devices to access these platforms. For instance, nearly 90\% traffic to Taobao is accounted for by its mobile app.

To promote the sales revenue of online merchants and also benefit the platform, most E-commerce Websites incorporate the sponsored search (SS) idea from traditional search engines. In SS, advertisers (merchants in our context) bid query keywords; when a user (customer) issues a query, auctions are triggered and the winning advertisements (products) are presented to the user. In the research field of SS, most previous work was focused on bidding optimization \cite{borgs2007dynamics, feldman2007budget, broder2011bid, zhao2018deep} and click prediction \cite{attenberg2009modeling, graepel2010web, wang2013psychological, gao2018attention}. In this work, we consider a novel problem for SS in the context of E-Commerce search: refining product advertisements (ads) in order to better attract query users to click them. Specifically, we aim to automatically add salient features of product ads which are the most attractive to the current query user, hoping to increase the click-through rate (CTR) of ads. We name such salient features as \textit{selling point keywords} (SPs) in this paper.

\begin{figure}
\centering 
\subfigure[]{
    \includegraphics[width=1.55in]{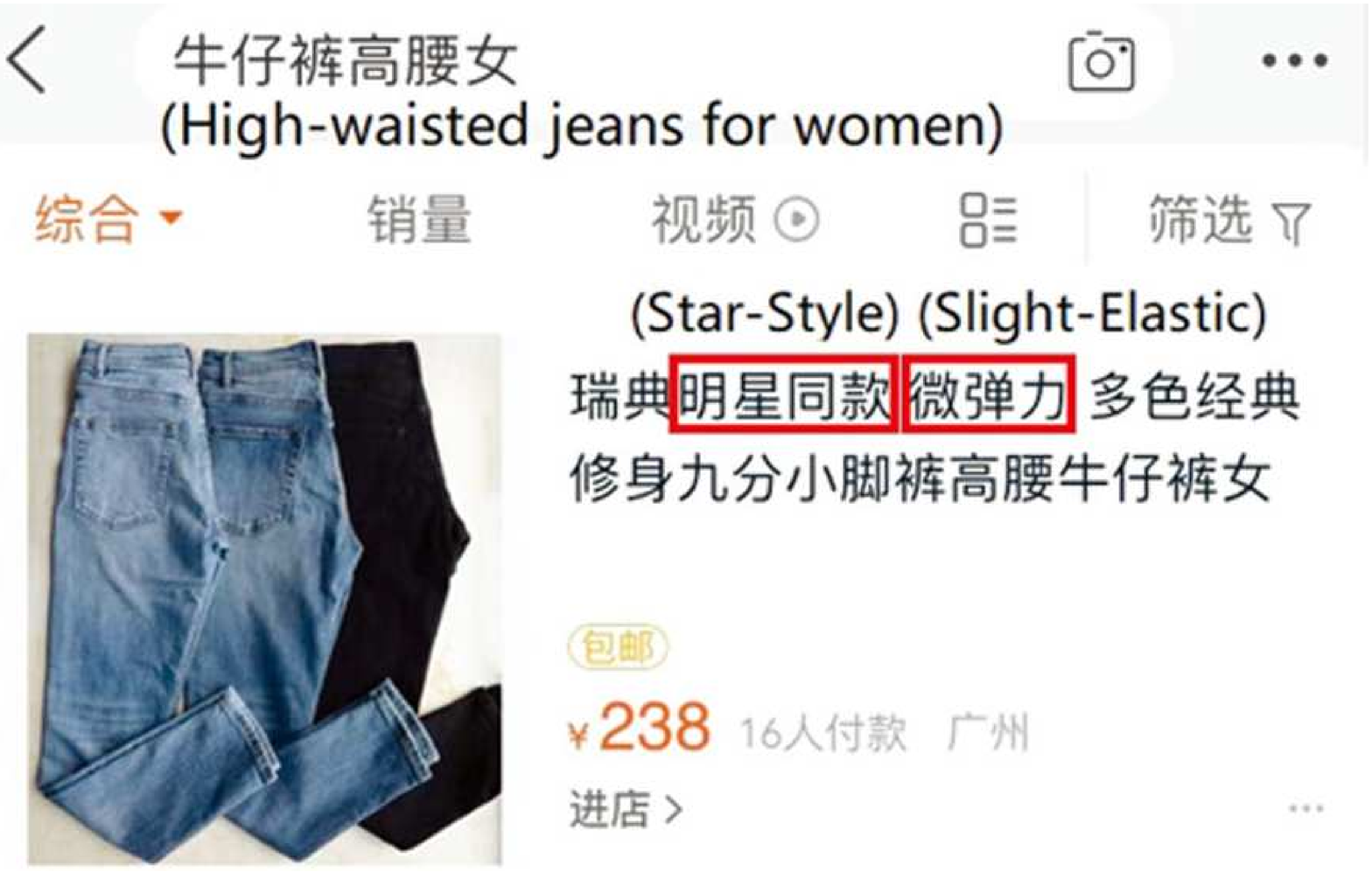}
    \label{sp-example.1}
}
\subfigure[]{
    \includegraphics[width=1.55in]{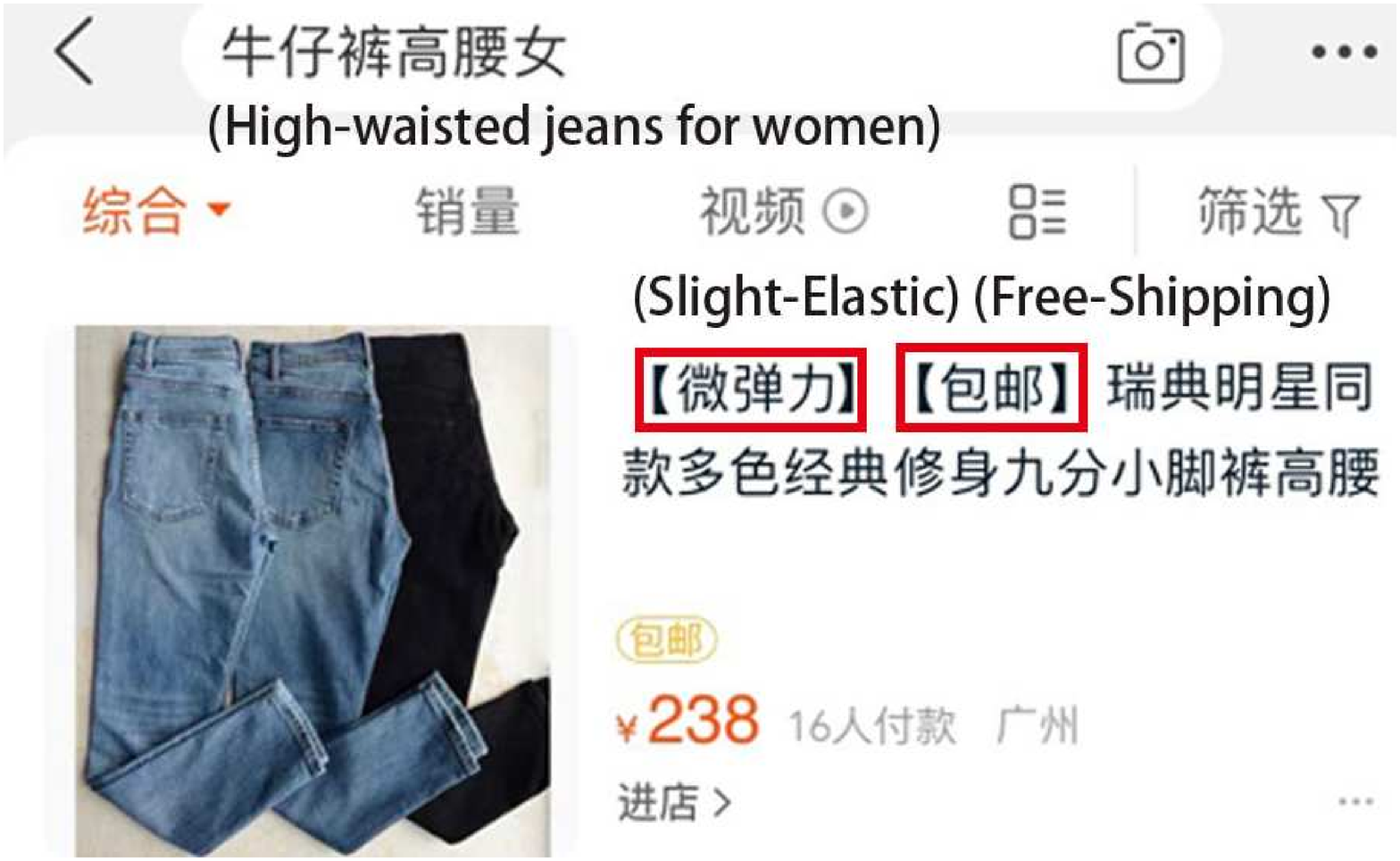}
    \label{sp-example.2}
}
\caption{An ad returned to a comfort-seeking user who is looking for ``High-waisted jeans for women'': (a) with the original title, and (b) with the refined title by our system. Texts in red bounding boxes are SPs. In the original title, SPs are hidden in the title, while in the refined one the most attractive SP is promoted to the front with emphasis.} 
\label{sp-example} 

\end{figure}

To our knowledge, no previous work has addressed this automatic ad refinement problem in the context of E-Commerce SS. In the context of general SS, some studies for click prediction have been concerned with analyzing the attractive factors of ads leading to clicks, e.g., specific text patterns such as ``official site'' \cite{wang2013psychological}, evidence types such as ``expert evidence'' and ``statistical evidence'' \cite{haans2013search}. These findings are important for helping advertisers to refine their ads in a general sense, but are orthogonal to our problem where we need to predict which SP best attracts the query user. Researchers have also proposed solutions for word attractiveness estimation \cite{kim2014advertiser,thomaidou2013automated,govindaraj2014modeling}. However, these methods all evaluated attractiveness without considering preference discrepancies among query users. In the E-Commerce context, personalization is very important since users can have quite different demands when looking for the same kind of products. Fig.~\ref{sp-example}(a) shows an example of the original product title in search results. We can see two examples SPs are hidden in the title. When looking for ``High-waisted jeans for women'', idolaters may prefer the first SP ``Star-Style'', while some other people may only be concerned about comfort and therefore we should make ``Slight-Elastic'' easily noticeable to them. Such inconspicuous SPs are not able to give users personalized experience, or even bypassed by them. In comparison, our system can put ``Slight-Elastic'' at the front with emphasis for users who like this feature (Fig.~\ref{sp-example}(b)). Providing personalized SP impressions to query users can not only benefit merchants and platforms by improving CTR, but also improve user search experience by showing their most concerned features to them. In Taobao, candidate SPs of an ad are mainly provided by the advertiser. Although Taobao also tries to automatically mine SPs, how to obtain candidate SPs is out of scope of this paper.

This automatic ad refinement problem is challenging due to the following reasons: (1) We must carefully manipulate the ad content so as to effectively attract users' attention without affecting their search experience. (2) Although user click data is abundant, it is difficult to obtain users' explicit feedback for their preference in product SPs. (3) As aforementioned, nowadays most traffic to E-Commerce platforms is from mobile devices. Due to space limitation on mobile devices, it is difficult to present many SPs to users.

\begin{figure}
\centering 
{\includegraphics[width=2.55in]{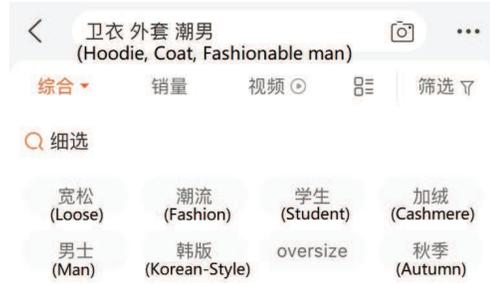}}
\caption{The search system of Taobao can provide specific feature keywords (SFs) related to the user query. Users can then click these SFs to narrow down the search space.} \label{sp-example2} 
\end{figure}

In this work, we are focused on the mobile setting and study how we can manipulate the ad titles by adding a few proper SPs to improve the CTR of ads. Firstly, we explore the impact of various exhibition schemes of SPs on the CTR performance. Secondly, we design a neural model for the task of personalized attractive SP prediction given a user and his/her submitted query. To alleviate the problem of lack of explicit user preference to SPs, we propose to employ the user click information on \textit{specific feature keywords} (SFs) as a surrogate to user explicit feedback for SP preference, and train the model accordingly. SFs are presented to a query user after he/she submits a query, to help narrow down the search space. Most E-Commerce Websites have this function. An example is shown in Fig.~\ref{sp-example2}. Like SPs, these SFs also represent the specific features of products. Nevertheless, the diversity and click data of SFs are still limited. On the other hand, users' historical CTR data can implicitly reflect their preference in SPs. For instance, if we observe that a user often clicks products with ``free shipping'' in the title, the user would probably favor products with such a feature. Hence, we employ CTR prediction as an auxiliary task to perform multi-task learning, trying to boost the performance of the main task (i.e., personalized attractive SP prediction). Finally, we further incorporate various additional features of users and queries into our model to better characterize them. Both large-scale offline evaluation and online A/B tests are systematically carried out in the Taobao platform to verify the effectiveness of these ideas.

A variant of our best model has already been deployed in Taobao, leading to a 2\% increase in revenue per thousand impressions and an opt-out rate of merchants less than 4\%. Readers can experience this new function in Taobao App.


\section{Effectiveness of Personalized Attraction Enhanced Sponsored Search}\label{sec:2}
In this section, we propose the first series of large-scale experiments and in-depth discussions to reveal the effectiveness of boosting the user attraction of an E-Commerce SS system by refining the ad content in search results. Using 120 million sponsored search impressions from Taobao, one of the largest E-Commerce platforms in China, online experiments and user studies are conducted to provide insights from users, merchants and the E-Commerce platform. 


To be specific, we refine the \textit{sponsored search results} (i.e., the returned records of ads) by automatically attaching personalized selling point keywords (SPs) to improve its attractiveness. In particular, experiments and analysis from this section aim to answer the following questions:
\begin{itemize}
\item \textit{Can the attractiveness of SS results be improved by refining their content?}
\item \textit{Is personalization essential for the attraction enhancement?}
\item \textit{How do different factors of content refinement, including the display format of the added SPs and the number of SPs, affect the attractiveness of SS results?}
\end{itemize}

Answering these questions is of great significance. On the one hand, they are related to the fundamental assumptions of our proposed attraction enhanced SS framework; On the other hand, the investigation of these questions brings the essential insight on how to build a more attractive and intuitive E-Commerce SS system. 


\subsection{Effectiveness of Personalized Refinement of Sponsored Search Results}
In this paper, we propose to refine the SS results and increase their attractiveness by adding personalized SPs in the front of the original SS results. Two assumptions need to be verified: 1.) adding personalized SPs helps ads to attract more user clicks than not adding; 2.) personalized exhibition of SPs helps to improve the attractiveness compared to non-personalized exhibition. 

Two online A/B tests are conducted independently to verify the two assumptions. For each test, search traffic from Taobao is equally split into the control and treatment groups. Click-through rate (CTR) is used as the metric to evaluate the attractiveness of the SS results. Furthermore, we also calculate the p-value according to Fisher's exact test \cite{upton1992fisher} to assess whether the results are significant. Three strategies are used to generate the SS results, including:

\begin{itemize}
\item \textbf{Title-based Sponsored Search result (TSS)} uses product titles provided by merchants to generate SS results and this is the traditional way to show ads, without any refinement.

\item \textbf{Personalized Selling Point keywords enhanced SS result (PSPSS)} adds personalized SPs in the front of ad title in the corresponding SS result. Two SPs most attractive to the target user are selected. Attractiveness is evaluated by our basic model introduced in detail in Section~\ref{sec:basic}.

\item \textbf{Non-personalized Selling Point keywords enhanced SS result (NSPSS)} picks the most popular SP keywords from an ad's candidate SP set according to the click log for that ad's category (thus capturing global preference of users for that category), and puts them in the front of the ad title in the corresponding SS result.

\end{itemize}

All the above-mentioned strategies share the same text-length constraint on the SS results. The part of the title that exceeds the text-length constraint will be cut off. 

\subsubsection{Effectiveness of Adding Personalized SPs}
As demonstrated in Experiment 1 from Table \ref{texp1}, compared with the traditional SS results without SPs (TSS), adding personalized SPs boosts the CTR of ads by 1.9\%. It is worth noting that the SS engine in Taobao is a well established system and usually a 0.3\% improvement on CTR is considered to be significant and the significance is also confirmed by the p-value. It strongly indicates adding SPs to SS results is a right direction for enhancing ad attractiveness. The effectiveness is also proved by our user study with over 95\% surveyed users preferring the enhanced version of SS results. It is worth noting the personalized SP prediction model used in PSPSS is just the basic version of our proposed models. After the initial online evaluation confirms our assumptions, more advanced models are designed and empirically tested with both online and offline experiments. Details will be reported in Sections~\ref{sec:mtl},~\ref{sec:augmented} and~\ref{sec:exp}.

There are two clear reasons behind the significant improvement. On the one hand, the added SPs reflect the products' most attractive features to the target users and inevitably help to attract users to click on the enhanced SS results. On the other hand, nowadays most of the users use their mobile devices for online shopping (90\% of Taobao's search traffic comes from the mobile-end). The display space for a SS result on a mobile device is usually very limited. For instance, in Taobao app, an SS result can only use up to 28 Chinese characters for its title and users have to make their decision with the information provided in the limited space. The personalized SPs added by our model usually reflect the features that a user cares most. This greatly increases the informativeness of the SS results and thus increases click-through rate. 

\subsubsection{Effectiveness of Personalization}
In our second A/B test, we investigate how much personalization helps to boost the attractiveness of SS results by comparing adding personalized SPs (PSPSS) vs. adding non-personalized SPs (NSPSS). As demonstrated in Experiment 2 from Table \ref{texp1}, a 0.55\% improvement is observed on CTR with the help of personalization. The p-value of 0.0009 also indicates a significant result. Experiment 2 confirms that adding personalized SPs is the right direction for our algorithm design.

The significant improvement is expected. Compared with non-personalized SPs generated in NSPSS, the personalization of PSPSS can help the search engine better target a user by demonstrating the features that he/she is the most interested in. This is also backed up by our user study: 82\% of the surveyed users point out that personalization can help them better locate the products they want.

\begin{table}[h]
    \centering
    \caption{Effectiveness of Personalized Refinement (CTR scores reflect relative CTR change of treatment over control).}
    \begin{tabular}{p{1cm}|p{1cm}|p{1.2cm}|p{0.8cm}|p{1cm}|p{1.5cm}}
        \hline
         Exp. ID&Control & Treatment & CTR & P-Value &Total Impressions \\
        \hline
         1&TSS & PSPSS & +1.9\% & 0.00001 & 20,500,000 \\
         \hline
         2&NSPSS & PSPSS &+0.55\% &0.0009 &40,874,000 \\
         \hline
    \end{tabular}

    \label{texp1}
\end{table}

\begin{table*}
\caption{The impact of SP emphasis.}
    \begin{tabular}{cccccc}
         \hline
         Experiment ID&Control&Treatment&CTR Change (Treatment Over Control)& P-Value&Total Impressions  \\
         \hline
         3&No Emphasis& With Emphasis & +0.39\% & 0.0167&20,500,000  \\
         \hline
    \end{tabular}
    
    \label{texp3}
\end{table*}

\begin{table*}
 \caption{The impact of the number of SPs.}
    \begin{tabular}{cccccc}
         \hline
         Experiment ID&Control&Treatment&CTR Change (Treatment Over Control)& P-Value&Total Impressions  \\
         \hline
         4&2 SPs added& 3 SPs added & +0.09\% &0.7347 &21,490,000
         \\
         5&2 SPs added& 1 SP added & -0.53\% &0.001 &20,200,000
         \\
         \hline
    \end{tabular}
    
    \label{texp4}
\end{table*}

\subsection{Impact of SP Exhibition Factors}
Besides the algorithm to pick the best personalized SPs (discussed in detail in the next sections), there are also some exhibition factors to be considered for SS results refinement. In particular: How do we emphasize the added SPs so as to attract query users? How many SPs should we add to a single SS result? These factors in fact proved to be very important for building an effective attraction enhanced E-Commerce SS system based on our large-scale A/B tests. 

\subsubsection{Impact of SP Emphasis}
In this part, we aim to investigate whether UI-based emphasis on the added SPs will make a difference to the attractiveness of SS results.

\begin{figure}
    \centering
    {\includegraphics[width=3.35in]{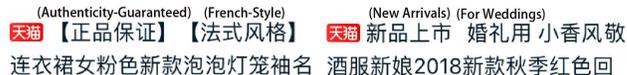}}
    \caption{The left case shows two SPs wrapped in bold box brackets which highlight them; the right case shows two SPs without emphasis.}
    \label{fempasis}
\end{figure}

To evaluate the impact of SP emphasis, we conduct an online A/B test to compare two types of SP display UI. The emphasized version comes with bold box brackets around each SP to highlight it and the non-emphasized version comes without the brackets (examples in Fig.~\ref{fempasis}). As displayed in Table \ref{texp3}, this type of emphasis helps to improve CTR by 0.39\%, with p-value=0.0167 (significant when significance level=0.05). We also tried out parentheses, normal box brackets and square brackets, but none of them works as good as the bold box brackets.

The explanation behind the scene can be interesting. The original SS result, which directly displays the title provided by the merchant, is usually lengthy and packed with merchant-provided keywords. Users can easily get swamped by the information and ignore the added SPs, as shown by the lower case of Fig.~\ref{fempasis}. The emphasis UI helps us grab users' attention to our proposed personalized SPs, which further enhances the attractiveness of SS results and consequently improves CTR.

\subsubsection{Impact of the Number of SPs}
Since we can add multiple SPs to a SS result, a natural question is how many SPs we should add. Does it follow "the more, the better" rule? Two online A/B tests are conducted to investigate the impact of adding different numbers of SPs. We investigate the CTR improvements of ``adding two SPs vs. adding three SPs'' (Experiment 4) and ``adding two SPs vs. adding one SPs'' (Experiment 5). The results are shown in Table~\ref{texp4}. Adding only one SP is significantly outperformed by adding two SPs. While the improvement of three SPs over two SPs is 0.09\%, its p-value shows no significance.

The results indicate only showing one SP is not reliable. One reason could be the imperfectness of the prediction model, i.e., failing to put the SP a user likes the most at the 1st position. Although we could improve the prediction model, showing two SPs seems to be more safe and do not hurt user experience. Regarding ``two SPs vs. three SPs'', there are two possible explanations. First, adding too many SPs tends to be overwhelming and distract a user from noticing the SP that truly attracts him. Secondly, the display space on mobile apps is usually very limited: each SS result on the Taobao app can only display 28 Chinese characters. Adding too many SPs will force a large portion of the original product title to be cut off and lead to negative user experience.

\begin{figure*}
\centering
{\includegraphics[height=2.1in,width=7in]{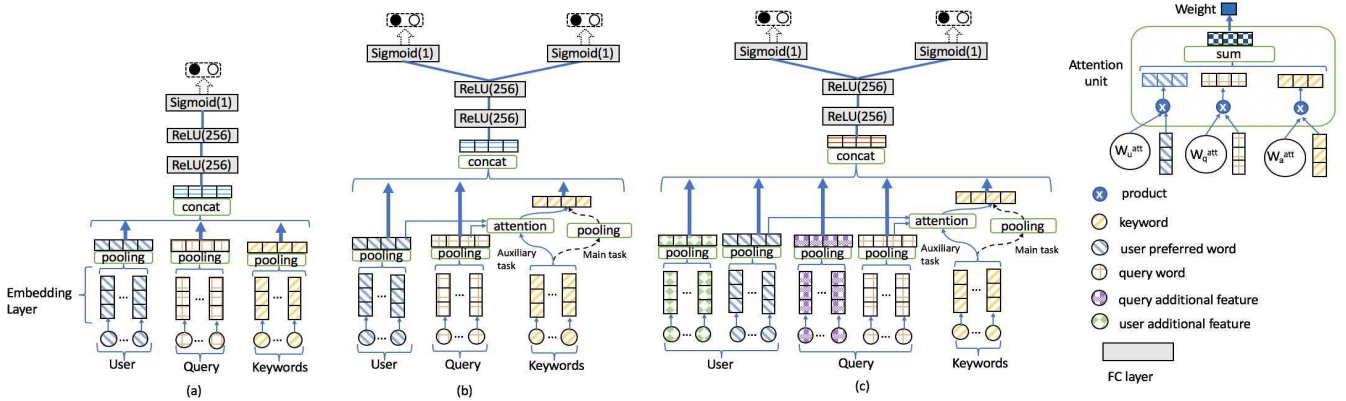}}
\caption{The three proposed models for personalized attractive SP prediction: (a) the basic model, (b) the multi-task model, and (c) the augmented model with additional features of uses and queries.}
\label{model}
\vspace{-4mm}
\end{figure*}

\section{The Basic Model}\label{sec:basic}
We have shown the importance of personalization for attractive selling point keyword (SP) prediction. In the next, we will detail the models we develop for personalized attractive SP prediction. We present firstly in this section our basic neural model for this task. This model is trained with users' click data on specific feature keywords (SFs) introduced in Section~\ref{sec:intro}.

\subsection{Common Notations \& Formulation}
We are given a set of users $\mathcal{U}$ and a set of ads $\mathcal{A}$. Let $\mathcal{W}$ denote the set of all keywords. Each ad $a \in \mathcal{A}$ has a set of SPs denoted as $\mathcal{T}_a=\{t_1, t_2,\dots,t_m\}$, where $t_i \in \mathcal{W}$. When a user $u \in \mathcal{U}$ submits a query $q=\{w_1, w_2,\dots, w_s\}$ ($w_i \in \mathcal{W}$), the search engine returns a set of ads. The goal is, for each returned ad $a$, to predict $p(t_i | u, q)$, the probability that each $t_i \in \mathcal{T}_a$ attracts $u$ given $q$. Then the SPs with the highest probabilities can be added to the SS result of $a$ in the search result list. Note the condition in $p(t_i | u, q)$ does not contain $a$. This is because given a submitted query, a user's preference in product features is deterministic and independent of specific ads.

In this paper, we estimate $p(t_i | u, q)$ via a neural model. Since we do not have users' explicit feedback for their preference in SPs, we employ users' click information on specific feature keywords (SFs) to train the model. Formally, we have a set of labeled SF click data $\mathcal{S}^{SF} = \{(u,q,v,y)\}$, where $v \in \mathcal{W}$ denotes a SF and $y$ is the binary label indicating whether $u$ clicks $v$\footnote{We will detail how to sample negative instances in the appendix}. The problem becomes, training a neural model for estimating $p(t_i | u, q)$ with supervision on $\mathcal{S}^{SF}$.


\subsection{Model Architecture \& Training}
The model structure of our basic model is depicted in Fig.~\ref{model}(a). At the input layer, we have three parts: user $u$, query $q$ and SF $v$. $q$ is a set of keywords and $v$ is also a keyword. As to the user $u$, we extract his/her recent click history for representation construction. Specifically, we use the frequent keywords in a user's click history to represent his/her long-term and short-term preferences (details can be found in the appendix). The advantage of this user representation scheme is two-fold: (1) modeling users' long-term and short-term interests conveyed by keywords can achieve proper personalization in estimating $p(t_i | u, q)$; (2) the ``out-of-sample'' issue can be naturally handled. That is, we can easily generalize the model to users unseen in the training stage. We can generate a user's representation as long as he/she has clicked a few products. 

$u$ and $q$ are modeled as multi-hot encoding vectors based on $\mathcal{W}$. The details of the remaining layers are as follows:
\begin{itemize}

\item \textbf{Embedding Layer}: The keyword set $\mathcal{W}$ is typically a very large set, resulting in very high dimensionality of multi-hot vectors. A popular way for deep learning models to reduce the dimensionality of multi-hot encoding vectors is to add embedding layers \cite{mikolov2013distributed}, which transform them into low dimensional dense vectors via an embedding table. Let $\mathbf{Em}(w)$ be a look-up function that returns the embedding vector for keyword $w$. The embedding layer transforms $u$, $q$ and $v$ into the corresponding embedding representations. For example, $q$ becomes $\{\mathbf{Em}(w_1), \mathbf{Em}(w_2),\dots, \mathbf{Em}(w_s)\}$.

\item \textbf{Pooling Layer \& Concatenation Layer}: The Embedding Layer transforms $u$ and $q$ into variable-size sets of vectors. To obtain fixed-length representations to facilitate further computation, we add a Pooling Layer above the Embedding Layer (average pooling is used in this work, e.g., for query: $\mathbf{e}_q = \frac{1}{s}\sum_{i=1}^{s} \mathbf{Em}(w_i)$). Note that a pooling operation is also placed over the SF embedding to make the model general. This is because a few SFs/SPs may contain multiple keywords. However, we still call them keywords for clarity and simplicity. After pooling, the Concatenation Layer simply synthesizes information from the three channels: $\mathbf{x} = [\mathbf{e}_u^T, \mathbf{e}_q^T, \mathbf{e}_v^T]^T$.

\item \textbf{Fully-connected Layers \& Output Layer}:
We then feed the obtained $\mathbf{x}$ through two Fully-connected (FC) Layers with the RELU nonlinear activation function \cite{lecun2015deep} to increase the expressiveness of the model. The final Output Layer is simply a logistic regression for label prediction.

\end{itemize}

The training protocol for the basic model is standard supervised training with the cross-entropy loss.

What we want to point out is that, this basic model is simple without in-depth techniques. However, we choose to stick to this simple design since: (1) the model should be able to generate real-time predictions when deployed in Taobao's search engine; (2) the major focus of this work is to investigate whether and how we can exploit user behaviors (click information on products and SFs) and rich features for personalized attractive SP prediction. We leave possible improvements for the model to future work.

\section{The Multi-task Model}\label{sec:mtl}
This section details how we use the CTR prediction task to boost the performance of the personalized attractive SP prediction task. We refer to them as \textit{auxiliary task} and \textit{main task}, respectively.

\subsection{Why using CTR prediction?}
Although users' click data on SFs can be regarded as providing direct supervision for the SP prediction task, the available information is still limited in the following two aspects: (1) the number of SF clicks is limited. In product search, users are more likely to click a returned product record than a shown SF; (2) the set of all SFs is limited and only covers a subset of the set of SPs.

In comparison to clicks on SFs, there are much more clicks on product records in product search and the product titles cover more diverse SPs\footnote{The Merchants always try to increase their products' visibility and CTR by adding exhaustive feature keywords in the title.}. Users' product click information can be readily obtained from the search engine log. Although product click data cannot provide direct supervision signals for the SP prediction task, they do imply to some extent how a user is attracted by the returned products in the context of the corresponding queries. For example, if we often observe that a user clicks products with the keyword ``Slight-Elastic'' in the title when searching for ``Jeans'', it is reasonable to believe the user likes jeans with slight elasticity as a feature. Hence, click data can be regarded as providing weak/implicit supervision for our main task. Though most of the time users are attracted by feature keywords in product titles, sometimes users may also be attracted by other factors of the returned product records, such as pictures. To reduce the impact of other factors, we collect only the click data for ads equipped with SP exhibition (generated by the basic model). When looking at these ads, users are more likely to be attracted by their titles due to the attractive presentation style (see Fig.~\ref{fempasis}). Formally, we define $\mathcal{S}^{C} = \{(u,q,a,y^c)\}$ to be a set of labeled user ad click data, where $a \in \mathcal{A}$ denotes an ad and $y^c$ is the binary label indicating whether $u$ clicks $a$.

\subsection{Model Architecture}
The multi-task model is shown in Fig.\ref{model}(b). As can be seen, we take a hard parameter sharing scheme \cite{ruder2017overview}. The overall structure is similar to the basic model. The only two differences are: (1) the output layers are separate for the two different tasks; (2) the input units for SPs (SFs) also accept ad titles, but we substitute an attention module for the pooling function at the pooling layer for the auxiliary task. The intuition for incorporating attention is that, since product titles typically contain much more keywords than SPs/SFs, an attention module can help distill the most important keywords explaining the corresponding click and also make the output distributions at the pooling layer more compatible between the two tasks. Let $\mathcal{D}_a = \{d_1, d_2, \dots, d_n\}$ denote the set of keywords in title $\mathcal{D}_a$ of ad $a$. We adopt a popular attention mechanism which is specified as follows
\begin{align}
    b_j &= \mathbf{z}^T \tanh( \mathbf{W}_u^{att} \mathbf{e}_u + \mathbf{W}_q^{att} \mathbf{e}_q + \mathbf{W}_a^{att} \mathbf{Em}(d_j) ) \nonumber \\
    \alpha_j &= \frac{\exp(b_j)}{\sum_{i=1}^n \exp(b_i)}, \quad \mathbf{e}_a = \sum_{j=1}^n \alpha_j \mathbf{Em}(d_j) \nonumber
\end{align}
Here we estimate the attention score of each word in $\mathcal{D}_a$ with respect to both $u$ and $q$, to obtain a personalized result. 

We believe that whether a user clicks an ad with SP exhibition is highly correlated with the attractiveness of the exhibited SPs to him/her. Hence, the two tasks are well correlated. By hard parameter sharing, useful knowledge from the auxiliary task could be transferred to the main task by the parameters for mapping users, queries and SPs/ad titles into high-level representations, thus facilitating the SP prediction problem. The training details of this model can be found in the appendix.

\section{The Augmented Model}\label{sec:augmented}
In this section, we present the augmented model equipped with both multi-task learning and additional features of the concerned user and query.

\subsection{Incorporating Features into the Model}
The augmented model is the same as the multi-task model, except for the input layer, where we add additional features for users and queries to better characterize them (green and purple nodes in Fig.~\ref{model}). In our framework, features are represented in a multi-group mutli-hot encoding form. Each group contains multiple discrete categorical features or bag-of-words (BoW) features that are semantically related. For example, suppose we have three features for users: \textit{gender}, \textit{occupation} and \textit{preferred brands}. The former two are discrete categorical features, while the last one is a BoW feature. These features form two groups: \{gender, occupation\} (user profile) and \{preferred brands\} (user preference). For a user with feature values [gender=male, occupation=doctor, prefer brand=\{Nike,Adidas\}], the corresponding multi-group mutli-hot encoding vector could be:
\begin{equation}
[~\{\underbrace{1,0}_{gender}~;~\underbrace{0,...,1,...,0}_{occupation}~\}~\{\underbrace{0,...,1,...,1...,0}_{preferred~ brands}\}~] \nonumber
\end{equation}
where we use ``\{\}'' as group delimiters and use ``;'' to separate features within the same group. This multi-group mutli-hot encoding vector is then fed to the embedding layer and the pooling layer to generate fixed-length vectors, as in the basic model. Here the pooling operation is performed within each group. 

\subsection{Features}\label{proposed_features}
Table \ref{Tab.feature} shows all the additional features used in our framework. In case a feature is numerical or continuous, it is discretized. We discuss these added features group by group. Detailed information of these features can be found in the appendix.

\vspace{2mm} \noindent \textbf{User profile information.} This group includes users' demographic features. Compared to using the long- and short-term interests of users only, these features further characterize users in a fine-grained level. Users with different demographic features may show different preferences in product features even if their interests are similar. For example, among people who like wearing jeans, young people may prefer the SP ``Star-style'' while old ones care more about comfort. Thus, demographic features could help the model better capture a user's preference in SPs.

\vspace{2mm} \noindent \textbf{User general preference.} This group of features encodes a user's preference regarding some important general aspects: product categories, product brands and whether the user likes discounts. These features could be useful for both tasks: for the auxiliary task, the general preference of a user accounts for the bias of user behaviors, which could better help the model correctly explain clicks so that more accurate representations could be transferred to the main task; for the main task, these features further enrich user preference and could help better evaluate personalized attractiveness. For instance, we could present brand-related (or discount-related) SPs to a user if we know the user has preferred brands (or likes discounts).

\vspace{2mm} \noindent \textbf{User consumption/activity level.} these features measure the consumption level and/or activity level of a user from different angles. Users with different consumption/activity levels may have different behavioral biases and tastes. For example, a user with high consumption level could care less about price but pay more attention on quality. Incorporating these features could also benefit both the two tasks.

\vspace{2mm} \noindent \textbf{Query category.} For queries, we add their category information (estimated by another module in the search engine) as a BoW feature to provide contextual information. The intuition is that, users generally pay attention to different features for products of different categories. For instance, people often concentrate on performance for PC, while for laptops weight is usually more important. If we only have a user's historical click information for the PC category, the basic and multi-task models may wrongly rank SPs related to performance higher when the user searches for laptops.

\begin{table}
\caption{A summarization of additional features used.}
\label{Tab.feature}
\begin{tabular}{|c|l|p{3cm}|}
\hline
Entity & Feature& Description \\
\hline
\multirow{14}*{User} &gender, age, occupation &\multirow{2}*{user profile information}\\
     &city, province & \\
     \cline{2-3}
     &preference for categories&\multirow{3}*{user general preference}\\
     &preference for brands&  \\
     &preference for discount&  \\
     \cline{2-3}
     &purchase level, VIP level&\multirow{3}*{\shortstack{user consumption/\\activity level}} \\
     &high consumption visitors&  \\
     &top class visitors&    \\
     \cline{2-3}
    \hline
    Query &category & query category\\
    \hline
\end{tabular}
\end{table}

\section{Experiments}\label{sec:exp}
Based on large-scale online experiments in Section~\ref{sec:2}, it has already been proved that adding personalized selling point keywords (SPs) to Sponsored Search (SS) results leads to more attractive SS results compared with not adding SPs or adding non-personalized SPs. In this section, both online and offline experiments are conducted to compare the personalized SP prediction models proposed in Section 3-5. We also analyze important factors that impact the performance of these models. In particular, we aim to answer the following questions: (1) How well do different versions of our proposed personalized SP prediction neural model work? (2) Will multi-task learning help to improve the performance? (3) Are the additional features introduced in Section~\ref{sec:augmented} useful? (4) How do different training strategies (see Section~\ref{training_methods} of the appendix) affect the performance of the multi-task model?

\subsection{Datasets}
In collaboration with Alibaba, our datasets are collected from Taobao, its C2C platform with over 500 million monthly active users. For training and offline evaluation purpose, we collect over 7 million users' click-through records for click information on specific feature keywords (called SF dataset) and on product search advertisements (called AD dataset). Statistics details of the collected SF and AD datasets are summarized in Table \ref{Table.dataSet}. The SF dataset is used to train our basic model and also serves as the input of the main task for our multi-task learning models. The AD dataset is used as the input of the auxiliary task of our multi-task learning models. It is worth noting that, as displayed in Table \ref{Table.dataSet}, we also calculate the percentage of users with over 10 clicks in the SF and AD datasets. AD dataset has 6 times of users with over 10 clicks compared with SF data, which makes AD a good auxiliary data source for our main task. Details of processing of the two datasets are in the appendix.

\begin{table}
    \centering
     \caption{The statistics of datasets.}
    \begin{tabular}{|c|c|c|}
    \hline
    & SF dataset & AD dataset  \\
    \hline
    total \# of clicks &47,356,924 & 290,336,555 \\
    \hline
    total \# of users & \multicolumn{2}{|c|}{7,946,738} \\
    \hline
    \% of users with over 10 clicks& 12.24\% & 76.13\%  \\
    \hline
    \end{tabular}
   
    \label{Table.dataSet}
\end{table}

\subsection{Compared Algorithms}\label{comparisonalgorithms}
Three algorithms corresponding to the three models proposed in the above are compared in this section: \textbf{The Basic Model} is proposed in Section~\ref{sec:basic} and is trained only using users' click data on specific feature keywords (SFs); \textbf{The Multi-task Model} is proposed in Section~\ref{sec:mtl} which uses both SF data and AD data at the same time; \textbf{The Augmented Model} is proposed in Section~\ref{sec:augmented} which extends the multi-task model by adding additional features.

The Area Under the receiver operating characteristic Curve (AUC) \cite{bradley1997use} is used as the evaluation metric for offline experiments and click-through rate (CTR) is used as the metric for online experiments. As in Section~\ref{sec:2}, p-value for Fisher's exact test is calculated to assess the significance of improvements in online experiments.




\subsection{Comparison Analysis}\label{algorithm_analysis}
We compare the three proposed models with both offline and online experiments. In the offline setting, the SF dataset is considered to be the ground truth representing users' preference on SPs and the three models are evaluated on the holdout test set. And in the online setting, we deploy the three models in Taobao's search engine and use large-scale online A/B tests to compare them. Totally over 90 million search impressions are collected in our A/B tests and in each A/B test, traffic is equally divided into the control and treatment groups. The results are reported in Tables \ref{table.methods_auc} and \ref{table.online_mtl} respectively. From Table \ref{table.online_mtl}, We can see both the augmented model and the multi-task model significantly outperform the basic model, with significance level 0.05. The augmented model shows the best performance. These results are consistent with the offline results.


These results are within our expectation. The multi-task model beats the basic model which is only trained on SF data. The reasons should be that, the volume of the SF data is limited and the set of SF keywords can only cover a subset of SP keywords. By introducing multi-task learning with CTR prediction for ads as an auxiliary task, extra useful information on users' implicit preference towards SPs is transferred to the personalized SP prediction task and thus improves its performance. 

The augmented model exceeding the multi-task model reveals another important insight in designing an effective SP prediction model: besides the model itself, the features used in the model also play an important part in improving its effectiveness. 


\subsection{Dissection of the Multi-task Model Results}
As mentioned in Section~\ref{algorithm_analysis}, both offline and online experiments confirm that multi-task learning helps to improve the performance of SP prediction by introducing auxiliary knowledge from a user's ad click history. In this section, we analyze the results of the multi-task model in detail to find out what kinds of users can benefit more from multi-task learning. 

We divide instances from the test set of SF dataset into four groups based on the corresponding users' numbers of clicks for main task and auxiliary task in the training data. That's to say, we try to find out whether a user has abundant information in the training data of main and auxiliary tasks will affect the performance gain of the multi-task model over the basic model. Due to the power-law effect, we choose the median of the number of user clicks in the main/auxiliary task training data as the threshold for dividing user. The results are summarized in Table \ref{table.action_auc}.

Based on the results, the highest performance gain is obtained for users with insufficient data in the main task but with abundant data in the auxiliary task (the first line). This makes sense because our proposed multi-task model aims to solve the data sparsity problem in the main task by transferring knowledge from the auxiliary task. Similarly, results from Table \ref{table.action_auc} also confirm that the lowest performance gain is observed for users who have abundant data in the main task but insufficient data in the auxiliary task. For them, we still can get a 2.99\% improvement. 


\subsection{Feature Ablation Study}
As confirmed in Section \ref{algorithm_analysis}, features play an important role in improving SP prediction performance. In Section \ref{proposed_features}, four groups of features were proposed to be added as additional features. In this section, we conduct an ablation study of those features. Specifically, we perform a series of offline experiments by adding these groups of features one at a time and evaluate the effectiveness of each group of features based on the gain of AUC. As shown in Table \ref{table.feature_auc}, all the four groups of features proposed in Section \ref{proposed_features} bring useful information for personalized SP prediction. When integrated with all of the four groups, our model achieves the best performance, indicating they are generally complementary to one another.


\subsection{Influence of Different Training Strategies}
Two training strategies, pre-training and alternative training, are introduced in Section~\ref{training_methods} of the appendix. In this section we compare these two training strategies in the offline experimental setting.

The results are presented in Table \ref{table.train_method}. We find that the pre-training strategy performs worse than the alternate training strategy. The main reason behind this could be that in alternate training, the main task interacts and optimizes together with the auxiliary task continuously, while in pre-training the auxiliary task only serves as a prior for the main task, without much interaction.

\subsection{Online System Deployment}
Serving over 500 million active users is not a trivial task. In deployment, we encountered some challenges. One of the main concern in designing our algorithm is to balance between complexity of the model and computational efficiency. Taobao has strict constraints on the response time of algorithms deployed in production. In order to make real-time predictions with low latency, we accelerated online serving of industrial deep networks through AVX (Advanced Vector Extensions) instruction set supported by Intel CPUs. Thanks to it, the computational performance increased by about 7 times. Based on a large-scale online analysis, our proposed framework on average can generate personalized SPs for hundreds of SS results for a single user in less than 5 milliseconds, which supports a swift user experience. Second, after the system being in production for a while, we received several complaints about bad SPs from merchants. In response, we recalled these bad cases for analysis. Then we designed artificial rules to post-process the automatically mined SPs, to make them intuitive and easy to understand (e.g., from ``old man'' to ``for old man'').

\begin{table}
    \centering
    \caption{Offline comparison results.}
    \begin{tabular}{cc}
      \hline
      Model&AUC \\
      \hline
      The Basic Model  & 0.5995 \\
      The Multi-Task Model  & 0.6215 (+3.67\%)\\
      The Augmented Model & 0.6261 (+4.44\%) \\
      \hline
    \end{tabular}
    
    \label{table.methods_auc}
\end{table}

\begin{table}[h]
    \centering
    \caption{Results for online comparisons.}
    \begin{tabular}{p{1.8cm}p{2.5cm}p{0.8cm}p{0.8cm}p{1.5cm}}
    \hline
     Control & Treatment  & CTR  & P-Value &Total Impressions\\
      \hline
     Basic Model&Multi-Task Model & +0.24\%& 0.0273 & 46,780,000  \\
     Basic Model&Augmented Model & +0.49\% & 0.0025 & 46,200,000  \\
      \hline
    \end{tabular}
    
    \label{table.online_mtl}
\end{table}

\begin{table}[h]
    \centering
    \caption{Performance of the multi-task model w.r.t different groups of users divided according to the numbers of clicks in the training data of main/auxiliary tasks.}
    \begin{tabular}{c|c|p{1.5cm}|p{2.5cm}}
        \hline
         Main Task& Auxiliary Task &AUC of Basic Model &AUC of Multi-task Model \\
        \hline
         clicks<=6& clicks>21 &0.6107 & 0.6341 (+3.83\%) \\
         \hline
         clicks<=6 & clicks<=21 &0.6075&0.6297 (+3.65\%) \\
         \hline
         clicks>6 &  clicks>21 &0.5906&0.6113 (+3.50\%)  \\
         \hline
         clicks>6 & clicks<=21 &0.5850&0.6025 (+2.99\%) \\
         \hline
    \end{tabular}
    
    \label{table.action_auc}
\end{table}

\begin{table}[h]
    \centering
    \caption{Ablation study for features used in the augmented model.}
    \begin{tabular}{cc}
        \hline
        Description & AUC (Gain) \\
        \hline
        No additional features & 0.6215  \\
        Only add user profile information & 0.6234 (+0.31\%)\\
        Only add user general preference& 0.6225 (+0.16\%)\\
        Only add user consumption/activity level& 0.6226 (+0.18\%)\\
        Only add query category& 0.6246 (+0.50\%) \\
        Add all additional features& 0.6261 (+0.74\%) \\
        \hline
    \end{tabular}
    \label{table.feature_auc}
\end{table}

\begin{table}[h]
    \centering
    \caption{AUCs of different training strategies for multi-task learning.}
    \begin{tabular}{cc}
      \hline
      Training strategies  & AUC \\
      \hline
        Alternate training & 0.6215\\
        Pre-training&0.6132\\
        \hline
    \end{tabular}
    
    \label{table.train_method}
\end{table}

\section{Related Work}
Sponsored search (SS) has achieved great industrial success. It is the major business model for most online search service providers. In academia, SS has also attracted a lot of attention from researchers in related fields. Typical research problems for SS include bidding optimization \cite{borgs2007dynamics, feldman2007budget, broder2011bid, zhao2018deep}, click prediction \cite{attenberg2009modeling, graepel2010web, wang2013psychological, gao2018attention}, keyword suggestion \cite{wu2009advertising, ravi2010automatic}, auction mechanism design \cite{gatti2012truthful}, etc. However, the majority of existing works were focused on the platform side or the advertiser side. Only a few research works have considered the attraction of ads to users. Wang \textit{et al.} \cite{wang2013psychological} studied why users clicked ads from the Psychological aspect. They used Maslow's desire theory to model user psychological desire in SS and automatically mined textual patterns in ads triggering users' desires. Haans \textit{et al.} \cite{haans2013search} empirically explored the impact of evidence type on CTR and conversion rate based on Google AdWords. Kim \textit{et al.} \cite{kim2014advertiser} developed an advertiser-centric click prediction approach for mining attractive words from ad texts which can then be suggested to the advertisers for ad refinement. Govindaraj \textit{et al.} \cite{govindaraj2014modeling} used the estimated word occurring probabilities in clicked documents to assess attractiveness. Our work is different from them in two aspects: (1) the above studies were focused on the general SS setting, while we are concerned with attractive SP prediction in E-Commerce SS. No existing methods can be applied to solve this problem. (2) More importantly, none of the above works considered personalization which is important in the E-Commerce setting. Thomaidou \textit{et al.} \cite{thomaidou2013automated} proposed an heuristic solution for attractive ad description generation in the E-Commerce setting. However, the proposed method is not applicable to our problem and it did not consider personalization either.

Our work is also related to Multi-task learning (MTL) \cite{caruana1997multitask}. MTL aims to share useful information among multiple tasks to improve the model performance for each task. The motivation of MTL is that human beings often apply the knowledge learned from previous tasks to help learn a new task. Hence, MTL can be regarded as a form of transfer learning \cite{pan2010survey}, and is also related to other areas in machine learning such as multi-label learning \cite{zhang2014review}. Readers can refer to some recent surveys \cite{zhang2017survey, ruder2017overview} for a complete view of MTL. Recently, deep learning based MTL methods have achieved promising performance \cite{ruder2017overview}. The general idea is to share model (sub-)structures among different tasks. Two popular sharing schemes are hard parameter sharing \cite{long2015learning} and soft parameter sharing \cite{duong2015low}. Hard sharing means multiple tasks share the same copy of model (sub-)structure; soft sharing keeps separate models for different tasks and uses regularization to encourage the parameters to be similar. Our multi-task model takes the hard sharing scheme, since we believe the auxiliary task (CTR prediction for ads equipped with SP exhibition) is well correlated with the main task (attractive SP prediction). That is, whether a user clicks an ad with SP exhibition is highly correlated with the attractiveness of the exhibited SPs to him/her. Our multi-task model is slightly different from the standard hard sharing model in that, (1) the attention module is exclusive for the auxiliary task; (2) the two tasks do not share the same input, i.e., $(u,q,v)$ vs. $(u,q,a)$. Hence, the multi-task model is a customized MTL model for our problem.

\section{Conclusion}
In this paper, we studied a novel problem for E-Commerce sponsored search: enhancing the attractiveness of SS results by adding personalized attractive selling point keywords to them. We systematically carried out online A/B tests in Taobao to verify the feasibility of this idea and also explored proper schemes for SP exhibition. In addition, we tried to attack the problem of training an effective model for this task from three aspects: (1) using users' click data on specific feature keywords as a surrogate of users' explicit preference on SPs for model training; (2) incorporating the CTR prediction task as an auxiliary task to perform multi-task learning, in order to boost the performance of personalized SP prediction; (3) incorporating additional features of users and queries to further improve the performance. Large-scale offline and online experiments confirmed the effectiveness of these ideas.

%

\begin{acks}
This research was supported by the National Natural Science Foundation of China (Grant Nos. 61672409, 61522206, 61876144, 61876145), the Science and Technology Plan Program in Shaanxi Province of China (Grant No. 2017KJXX-80), the Fundamental Research Funds for the Central Universities (Grant Nos. JB190301, JB190305), and Alibaba-Zhejiang University Joint Research Institute of Frontier Technologies. The authors would like to thank the colleagues of Alimama for their supports, including Jinping Gou, Changyou Xu, Xinyang Guo, Sheng Xu, Jing Chen, Qian Wang and Ju Huang. 



\end{acks}

%
\bibliographystyle{ACM-Reference-Format}
\bibliography{main.bib}


\begin{thebibliography}{28}


\ifx \showCODEN    \undefined \def \showCODEN     #1{\unskip}     \fi
\ifx \showDOI      \undefined \def \showDOI       #1{#1}\fi
\ifx \showISBNx    \undefined \def \showISBNx     #1{\unskip}     \fi
\ifx \showISBNxiii \undefined \def \showISBNxiii  #1{\unskip}     \fi
\ifx \showISSN     \undefined \def \showISSN      #1{\unskip}     \fi
\ifx \showLCCN     \undefined \def \showLCCN      #1{\unskip}     \fi
\ifx \shownote     \undefined \def \shownote      #1{#1}          \fi
\ifx \showarticletitle \undefined \def \showarticletitle #1{#1}   \fi
\ifx \showURL      \undefined \def \showURL       {\relax}        \fi
\providecommand\bibfield[2]{#2}
\providecommand\bibinfo[2]{#2}
\providecommand\natexlab[1]{#1}
\providecommand\showeprint[2][]{arXiv:#2}

\bibitem[\protect\citeauthoryear{Attenberg, Pandey, and Suel}{Attenberg
  et~al\mbox{.}}{2009}]%
        {attenberg2009modeling}
\bibfield{author}{\bibinfo{person}{Josh Attenberg}, \bibinfo{person}{Sandeep
  Pandey}, {and} \bibinfo{person}{Torsten Suel}.}
  \bibinfo{year}{2009}\natexlab{}.
\newblock \showarticletitle{Modeling and predicting user behavior in sponsored
  search}. In \bibinfo{booktitle}{\emph{SIGKDD}}. ACM,
  \bibinfo{pages}{1067--1076}.
\newblock


\bibitem[\protect\citeauthoryear{Borgs, Chayes, Immorlica, Jain, Etesami, and
  Mahdian}{Borgs et~al\mbox{.}}{2007}]%
        {borgs2007dynamics}
\bibfield{author}{\bibinfo{person}{Christian Borgs}, \bibinfo{person}{Jennifer
  Chayes}, \bibinfo{person}{Nicole Immorlica}, \bibinfo{person}{Kamal Jain},
  \bibinfo{person}{Omid Etesami}, {and} \bibinfo{person}{Mohammad Mahdian}.}
  \bibinfo{year}{2007}\natexlab{}.
\newblock \showarticletitle{Dynamics of bid optimization in online
  advertisement auctions}. In \bibinfo{booktitle}{\emph{WWW}}. ACM,
  \bibinfo{pages}{531--540}.
\newblock


\bibitem[\protect\citeauthoryear{Bradley}{Bradley}{1997}]%
        {bradley1997use}
\bibfield{author}{\bibinfo{person}{Andrew~P Bradley}.}
  \bibinfo{year}{1997}\natexlab{}.
\newblock \showarticletitle{The use of the area under the ROC curve in the
  evaluation of machine learning algorithms}.
\newblock \bibinfo{journal}{\emph{Pattern recognition}} \bibinfo{volume}{30},
  \bibinfo{number}{7} (\bibinfo{year}{1997}), \bibinfo{pages}{1145--1159}.
\newblock


\bibitem[\protect\citeauthoryear{Broder, Gabrilovich, Josifovski, Mavromatis,
  and Smola}{Broder et~al\mbox{.}}{2011}]%
        {broder2011bid}
\bibfield{author}{\bibinfo{person}{Andrei Broder}, \bibinfo{person}{Evgeniy
  Gabrilovich}, \bibinfo{person}{Vanja Josifovski}, \bibinfo{person}{George
  Mavromatis}, {and} \bibinfo{person}{Alex Smola}.}
  \bibinfo{year}{2011}\natexlab{}.
\newblock \showarticletitle{Bid generation for advanced match in sponsored
  search}. In \bibinfo{booktitle}{\emph{WSDM}}. ACM, \bibinfo{pages}{515--524}.
\newblock


\bibitem[\protect\citeauthoryear{Caruana}{Caruana}{1997}]%
        {caruana1997multitask}
\bibfield{author}{\bibinfo{person}{Rich Caruana}.}
  \bibinfo{year}{1997}\natexlab{}.
\newblock \showarticletitle{Multitask learning}.
\newblock \bibinfo{journal}{\emph{Machine learning}} \bibinfo{volume}{28},
  \bibinfo{number}{1} (\bibinfo{year}{1997}), \bibinfo{pages}{41--75}.
\newblock


\bibitem[\protect\citeauthoryear{Duchi, Hazan, and Singer}{Duchi
  et~al\mbox{.}}{2011}]%
        {duchi2011adaptive}
\bibfield{author}{\bibinfo{person}{John Duchi}, \bibinfo{person}{Elad Hazan},
  {and} \bibinfo{person}{Yoram Singer}.} \bibinfo{year}{2011}\natexlab{}.
\newblock \showarticletitle{Adaptive subgradient methods for online learning
  and stochastic optimization}.
\newblock \bibinfo{journal}{\emph{JMLR}} \bibinfo{volume}{12},
  \bibinfo{number}{Jul} (\bibinfo{year}{2011}), \bibinfo{pages}{2121--2159}.
\newblock


\bibitem[\protect\citeauthoryear{Duong, Cohn, Bird, and Cook}{Duong
  et~al\mbox{.}}{2015}]%
        {duong2015low}
\bibfield{author}{\bibinfo{person}{Long Duong}, \bibinfo{person}{Trevor Cohn},
  \bibinfo{person}{Steven Bird}, {and} \bibinfo{person}{Paul Cook}.}
  \bibinfo{year}{2015}\natexlab{}.
\newblock \showarticletitle{Low resource dependency parsing: Cross-lingual
  parameter sharing in a neural network parser}. In
  \bibinfo{booktitle}{\emph{ACL-IJCNLP}}, Vol.~\bibinfo{volume}{2}.
  \bibinfo{pages}{845--850}.
\newblock


\bibitem[\protect\citeauthoryear{Feldman, Muthukrishnan, Pal, and
  Stein}{Feldman et~al\mbox{.}}{2007}]%
        {feldman2007budget}
\bibfield{author}{\bibinfo{person}{Jon Feldman}, \bibinfo{person}{S
  Muthukrishnan}, \bibinfo{person}{Martin Pal}, {and} \bibinfo{person}{Cliff
  Stein}.} \bibinfo{year}{2007}\natexlab{}.
\newblock \showarticletitle{Budget optimization in search-based advertising
  auctions}. In \bibinfo{booktitle}{\emph{ACM EC}}. ACM,
  \bibinfo{pages}{40--49}.
\newblock


\bibitem[\protect\citeauthoryear{Gao, Kong, Lu, Bai, and Yang}{Gao
  et~al\mbox{.}}{2018}]%
        {gao2018attention}
\bibfield{author}{\bibinfo{person}{Hongchang Gao}, \bibinfo{person}{Deguang
  Kong}, \bibinfo{person}{Miao Lu}, \bibinfo{person}{Xiao Bai}, {and}
  \bibinfo{person}{Jian Yang}.} \bibinfo{year}{2018}\natexlab{}.
\newblock \showarticletitle{Attention Convolutional Neural Network for
  Advertiser-level Click-through Rate Forecasting}. In
  \bibinfo{booktitle}{\emph{WWW}}. International World Wide Web Conferences
  Steering Committee, \bibinfo{pages}{1855--1864}.
\newblock


\bibitem[\protect\citeauthoryear{Gatti, Lazaric, and Trov{\`o}}{Gatti
  et~al\mbox{.}}{2012}]%
        {gatti2012truthful}
\bibfield{author}{\bibinfo{person}{Nicola Gatti}, \bibinfo{person}{Alessandro
  Lazaric}, {and} \bibinfo{person}{Francesco Trov{\`o}}.}
  \bibinfo{year}{2012}\natexlab{}.
\newblock \showarticletitle{A truthful learning mechanism for multi-slot
  sponsored search auctions with externalities}. In
  \bibinfo{booktitle}{\emph{AAMAS}}. International Foundation for Autonomous
  Agents and Multiagent Systems, \bibinfo{pages}{1325--1326}.
\newblock


\bibitem[\protect\citeauthoryear{Govindaraj, Wang, and Vishwanathan}{Govindaraj
  et~al\mbox{.}}{2014}]%
        {govindaraj2014modeling}
\bibfield{author}{\bibinfo{person}{Dinesh Govindaraj}, \bibinfo{person}{Tao
  Wang}, {and} \bibinfo{person}{SVN Vishwanathan}.}
  \bibinfo{year}{2014}\natexlab{}.
\newblock \showarticletitle{Modeling attractiveness and multiple clicks in
  sponsored search results}.
\newblock \bibinfo{journal}{\emph{arXiv preprint arXiv:1401.0255}}
  (\bibinfo{year}{2014}).
\newblock


\bibitem[\protect\citeauthoryear{Graepel, Candela, Borchert, and
  Herbrich}{Graepel et~al\mbox{.}}{2010}]%
        {graepel2010web}
\bibfield{author}{\bibinfo{person}{Thore Graepel},
  \bibinfo{person}{Joaquin~Qui{\~n}onero Candela}, \bibinfo{person}{Thomas
  Borchert}, {and} \bibinfo{person}{Ralf Herbrich}.}
  \bibinfo{year}{2010}\natexlab{}.
\newblock \showarticletitle{Web-scale Bayesian click-through rate prediction
  for sponsored search advertising in Microsoft's bing search engine}. In
  \bibinfo{booktitle}{\emph{ICML}}. Omnipress, \bibinfo{pages}{13--20}.
\newblock


\bibitem[\protect\citeauthoryear{Group}{Group}{2018}]%
        {ali2018report}
\bibfield{author}{\bibinfo{person}{Alibaba Group}.}
  \bibinfo{year}{2018}\natexlab{}.
\newblock \bibinfo{title}{Alibaba Group Announces March Quarter 2018 Results
  and Full Fiscal Year 2018 Results}.
\newblock
  \bibinfo{howpublished}{\url{https://www.alibabagroup.com/en/news/press_pdf/p180504.pdf}}.
\newblock


\bibitem[\protect\citeauthoryear{Haans, Raassens, and van Hout}{Haans
  et~al\mbox{.}}{2013}]%
        {haans2013search}
\bibfield{author}{\bibinfo{person}{Hans Haans}, \bibinfo{person}{N{\'e}omie
  Raassens}, {and} \bibinfo{person}{Roel van Hout}.}
  \bibinfo{year}{2013}\natexlab{}.
\newblock \showarticletitle{Search engine advertisements: The impact of
  advertising statements on click-through and conversion rates}.
\newblock \bibinfo{journal}{\emph{Marketing Letters}} \bibinfo{volume}{24},
  \bibinfo{number}{2} (\bibinfo{year}{2013}), \bibinfo{pages}{151--163}.
\newblock


\bibitem[\protect\citeauthoryear{Kim, Qin, Liu, and Yu}{Kim
  et~al\mbox{.}}{2014}]%
        {kim2014advertiser}
\bibfield{author}{\bibinfo{person}{Sungchul Kim}, \bibinfo{person}{Tao Qin},
  \bibinfo{person}{Tie-Yan Liu}, {and} \bibinfo{person}{Hwanjo Yu}.}
  \bibinfo{year}{2014}\natexlab{}.
\newblock \showarticletitle{Advertiser-centric approach to understand user
  click behavior in sponsored search}.
\newblock \bibinfo{journal}{\emph{Information Sciences}}  \bibinfo{volume}{276}
  (\bibinfo{year}{2014}), \bibinfo{pages}{242--254}.
\newblock


\bibitem[\protect\citeauthoryear{LeCun, Bengio, and Hinton}{LeCun
  et~al\mbox{.}}{2015}]%
        {lecun2015deep}
\bibfield{author}{\bibinfo{person}{Yann LeCun}, \bibinfo{person}{Yoshua
  Bengio}, {and} \bibinfo{person}{Geoffrey Hinton}.}
  \bibinfo{year}{2015}\natexlab{}.
\newblock \showarticletitle{Deep learning}.
\newblock \bibinfo{journal}{\emph{Nature}} \bibinfo{volume}{521},
  \bibinfo{number}{7553} (\bibinfo{year}{2015}), \bibinfo{pages}{436}.
\newblock


\bibitem[\protect\citeauthoryear{Long and Wang}{Long and Wang}{2015}]%
        {long2015learning}
\bibfield{author}{\bibinfo{person}{Mingsheng Long} {and}
  \bibinfo{person}{Jianmin Wang}.} \bibinfo{year}{2015}\natexlab{}.
\newblock \showarticletitle{Learning multiple tasks with deep relationship
  networks}.
\newblock \bibinfo{journal}{\emph{CoRR, abs/1506.02117}}  \bibinfo{volume}{3}
  (\bibinfo{year}{2015}).
\newblock


\bibitem[\protect\citeauthoryear{Mikolov, Sutskever, Chen, Corrado, and
  Dean}{Mikolov et~al\mbox{.}}{2013}]%
        {mikolov2013distributed}
\bibfield{author}{\bibinfo{person}{Tomas Mikolov}, \bibinfo{person}{Ilya
  Sutskever}, \bibinfo{person}{Kai Chen}, \bibinfo{person}{Greg~S Corrado},
  {and} \bibinfo{person}{Jeff Dean}.} \bibinfo{year}{2013}\natexlab{}.
\newblock \showarticletitle{Distributed representations of words and phrases
  and their compositionality}. In \bibinfo{booktitle}{\emph{NIPS}}.
  \bibinfo{pages}{3111--3119}.
\newblock


\bibitem[\protect\citeauthoryear{Pan, Yang, et~al\mbox{.}}{Pan
  et~al\mbox{.}}{2010}]%
        {pan2010survey}
\bibfield{author}{\bibinfo{person}{Sinno~Jialin Pan}, \bibinfo{person}{Qiang
  Yang}, {et~al\mbox{.}}} \bibinfo{year}{2010}\natexlab{}.
\newblock \showarticletitle{A survey on transfer learning}.
\newblock \bibinfo{journal}{\emph{IEEE TKDE}} \bibinfo{volume}{22},
  \bibinfo{number}{10} (\bibinfo{year}{2010}), \bibinfo{pages}{1345--1359}.
\newblock


\bibitem[\protect\citeauthoryear{Ravi, Broder, Gabrilovich, Josifovski, Pandey,
  and Pang}{Ravi et~al\mbox{.}}{2010}]%
        {ravi2010automatic}
\bibfield{author}{\bibinfo{person}{Sujith Ravi}, \bibinfo{person}{Andrei
  Broder}, \bibinfo{person}{Evgeniy Gabrilovich}, \bibinfo{person}{Vanja
  Josifovski}, \bibinfo{person}{Sandeep Pandey}, {and} \bibinfo{person}{Bo
  Pang}.} \bibinfo{year}{2010}\natexlab{}.
\newblock \showarticletitle{Automatic generation of bid phrases for online
  advertising}. In \bibinfo{booktitle}{\emph{WSDM}}. ACM,
  \bibinfo{pages}{341--350}.
\newblock


\bibitem[\protect\citeauthoryear{Ruder}{Ruder}{2017}]%
        {ruder2017overview}
\bibfield{author}{\bibinfo{person}{Sebastian Ruder}.}
  \bibinfo{year}{2017}\natexlab{}.
\newblock \showarticletitle{An overview of multi-task learning in deep neural
  networks}.
\newblock \bibinfo{journal}{\emph{arXiv preprint arXiv:1706.05098}}
  (\bibinfo{year}{2017}).
\newblock


\bibitem[\protect\citeauthoryear{Thomaidou, Lourentzou, Katsivelis-Perakis, and
  Vazirgiannis}{Thomaidou et~al\mbox{.}}{2013}]%
        {thomaidou2013automated}
\bibfield{author}{\bibinfo{person}{Stamatina Thomaidou},
  \bibinfo{person}{Ismini Lourentzou}, \bibinfo{person}{Panagiotis
  Katsivelis-Perakis}, {and} \bibinfo{person}{Michalis Vazirgiannis}.}
  \bibinfo{year}{2013}\natexlab{}.
\newblock \showarticletitle{Automated snippet generation for online
  advertising}. In \bibinfo{booktitle}{\emph{CIKM}}. ACM,
  \bibinfo{pages}{1841--1844}.
\newblock


\bibitem[\protect\citeauthoryear{Upton}{Upton}{1992}]%
        {upton1992fisher}
\bibfield{author}{\bibinfo{person}{Graham~JG Upton}.}
  \bibinfo{year}{1992}\natexlab{}.
\newblock \showarticletitle{Fisher's exact test}.
\newblock \bibinfo{journal}{\emph{Journal of the Royal Statistical Society.
  Series A (Statistics in Society)}} (\bibinfo{year}{1992}),
  \bibinfo{pages}{395--402}.
\newblock


\bibitem[\protect\citeauthoryear{Wang, Bian, Liu, Zhang, and Liu}{Wang
  et~al\mbox{.}}{2013}]%
        {wang2013psychological}
\bibfield{author}{\bibinfo{person}{Taifeng Wang}, \bibinfo{person}{Jiang Bian},
  \bibinfo{person}{Shusen Liu}, \bibinfo{person}{Yuyu Zhang}, {and}
  \bibinfo{person}{Tie-Yan Liu}.} \bibinfo{year}{2013}\natexlab{}.
\newblock \showarticletitle{Psychological advertising: exploring user
  psychology for click prediction in sponsored search}. In
  \bibinfo{booktitle}{\emph{SIGKDD}}. ACM, \bibinfo{pages}{563--571}.
\newblock


\bibitem[\protect\citeauthoryear{Wu, Qiu, He, Shi, Qu, Shen, Bu, and Chen}{Wu
  et~al\mbox{.}}{2009}]%
        {wu2009advertising}
\bibfield{author}{\bibinfo{person}{Hao Wu}, \bibinfo{person}{Guang Qiu},
  \bibinfo{person}{Xiaofei He}, \bibinfo{person}{Yuan Shi},
  \bibinfo{person}{Mingcheng Qu}, \bibinfo{person}{Jing Shen},
  \bibinfo{person}{Jiajun Bu}, {and} \bibinfo{person}{Chun Chen}.}
  \bibinfo{year}{2009}\natexlab{}.
\newblock \showarticletitle{Advertising keyword generation using active
  learning}. In \bibinfo{booktitle}{\emph{WWW}}. ACM,
  \bibinfo{pages}{1095--1096}.
\newblock


\bibitem[\protect\citeauthoryear{Zhang and Zhou}{Zhang and Zhou}{2014}]%
        {zhang2014review}
\bibfield{author}{\bibinfo{person}{Min-Ling Zhang} {and}
  \bibinfo{person}{Zhi-Hua Zhou}.} \bibinfo{year}{2014}\natexlab{}.
\newblock \showarticletitle{A review on multi-label learning algorithms}.
\newblock \bibinfo{journal}{\emph{IEEE TKDE}} \bibinfo{volume}{26},
  \bibinfo{number}{8} (\bibinfo{year}{2014}), \bibinfo{pages}{1819--1837}.
\newblock


\bibitem[\protect\citeauthoryear{Zhang and Yang}{Zhang and Yang}{2017}]%
        {zhang2017survey}
\bibfield{author}{\bibinfo{person}{Yu Zhang} {and} \bibinfo{person}{Qiang
  Yang}.} \bibinfo{year}{2017}\natexlab{}.
\newblock \showarticletitle{A survey on multi-task learning}.
\newblock \bibinfo{journal}{\emph{arXiv preprint arXiv:1707.08114}}
  (\bibinfo{year}{2017}).
\newblock


\bibitem[\protect\citeauthoryear{Zhao, Qiu, Guan, Zhao, and He}{Zhao
  et~al\mbox{.}}{2018}]%
        {zhao2018deep}
\bibfield{author}{\bibinfo{person}{Jun Zhao}, \bibinfo{person}{Guang Qiu},
  \bibinfo{person}{Ziyu Guan}, \bibinfo{person}{Wei Zhao}, {and}
  \bibinfo{person}{Xiaofei He}.} \bibinfo{year}{2018}\natexlab{}.
\newblock \showarticletitle{Deep Reinforcement Learning for Sponsored Search
  Real-time Bidding}.
\newblock \bibinfo{journal}{\emph{arXiv preprint arXiv:1803.00259}}
  (\bibinfo{year}{2018}).
\newblock


\end{thebibliography}

\newpage
%

\appendix

\section{Reproducibility}
In this appendix, we provide details of our implementation and experimental setup to help reproduce the findings in this work. We cannot release the codes and datasets due to business secret and privacy issues. However, the proposed models are rather standard without sophisticated techniques, so we believe it is easy to re-implement them with the information here.

\section{Training Details of the Multi-task Model}\label{training_methods}
The objective functions of the two tasks (personalized SP prediction and CTR prediction) are both cross-entropy loss:
\begin{equation}\label{eq.6.}
\mathcal{L}_{aux}= - \frac{1}{|\mathcal{S}^C|}\sum [y^c \log (y^c)'+ (1-y^c)\log(1-(y^c)')]
\end{equation}
\begin{equation}\label{eq.7.}
\mathcal{L}_{main}= - \frac{1}{|\mathcal{S}^{SF}|}\sum [y \log y' + (1-y)log(1-y')]
\end{equation}
where $\mathcal{L}_{aux}$ and $\mathcal{L}_{main}$ denote the loss functions for the auxiliary task and the main task, respectively, and $y'$/$(y^c)'$ represents the predicted label for an instance in $\mathcal{S}^{SF}$/$\mathcal{S}^C$.

The training of the model is slightly different from the classic multi-task models for which joint training can be performed. By joint training, we mean the classic multi-task models can do model updating jointly for different tasks based on the same mini-batch of instances. This is because the tasks share the same input. I.e., the same instance generates different outputs for different tasks. However, our two tasks do not share the same input: $(u,q,v)$ vs. $(u,q,a)$. For one instance from either $\mathcal{S}^{SF}$ or $\mathcal{S}^{C}$, we cannot do updating for the two tasks simultaneously. Hence, we explore two training strategies, \textit{alternate training} and \textit{pre-training}, for model training.


The alternate training algorithm is shown in Algorithm \ref{alg.1.}. We use $\Theta_1$/$\Theta_2$ to denote the set of parameters related to the auxiliary/main task. The iterative procedure processes the training data by sampling mini-batches. In each iteration, the main task or the auxiliary task is selected randomly, with a probability proportion of $1:k$. Then the related parameters are updated by picking a mini-batch from the corresponding dataset, i.e. $\mathcal{S}^{SF}$ or $\mathcal{S}^C$. We set the probability of choosing the auxiliary task $k$ times of that for the main task. The reason is that $\mathcal{S}^C$ is typically larger than $\mathcal{S}^{SF}$. $k$ is set so that the two datasets are processed approximately the same number of times when the algorithm stops. This assures the model is sufficiently trained on both datasets.

\begin{algorithm}[h]
\caption{Alternate Training}
\label{alg.1.}
Initialize model parameters randomly. $\Theta_1$ is the set of parameters related to the auxiliary task and $\Theta_2$ is the set of parameters related to the main task \\
\For{iteration in $1, 2, \dots$}{
    Choose the main task or the auxiliary task with a probability proportion of $1 : k$\\
    \If{task is auxiliary}{
        Pick a mini-batch from $\mathcal{S}^{C}$\\
        Compute loss $\mathcal{L}_{aux}(\Theta_1)$ according to Eq.(\ref{eq.6.})\\
        Compute gradient: $\nabla \mathcal{L}_{aux}(\Theta_1)$\\
        Update model: $\Theta_1=\Theta_1-\varepsilon\nabla \mathcal{L}_{aux}(\Theta_1)$
    }\Else{
        Pick a mini-batch from $\mathcal{S}^{SF}$\\
        Compute loss $\mathcal{L}_{main}(\Theta_2)$ according to Eq.(\ref{eq.7.})\\
        Compute gradient: $\nabla \mathcal{L}_{main}(\Theta_2)$\\
        Update model: $\Theta_2=\Theta_2-\varepsilon\nabla \mathcal{L}_{main}(\Theta_2)$
    }
}
\end{algorithm}

The pre-training strategy intrinsically performs transfer learning from the auxiliary task to the main task. As shown in Algorithm \ref{alg.2.}, we first train the model with respect to the auxiliary task using $\mathcal{S}^C$, to obtain learned $\Theta_1$. Then we incorporate these learned parameters (except the output layer and the attention module) as a prior and further train the model with respect to the main task. We will investigate the two training schemes in the experiments.

\begin{algorithm}[h]
\caption{Pre-training}
\label{alg.2.}
Initialize model parameters randomly. $\Theta_1$ is the set of parameters related to the auxiliary task \\
\For{iteration in $1, 2, \dots$}{
    Pick a mini-batch from $\mathcal{S}^{C}$\\
    Compute loss $\mathcal{L}_{aux}(\Theta_1)$ according to Eq.(\ref{eq.6.})\\
    Compute gradient: $\nabla \mathcal{L}_{aux}(\Theta_1)$\\
    Update model: $\Theta_1=\Theta_1-\varepsilon\nabla \mathcal{L}_{aux}(\Theta_1)$
}
Initialize shared model parameters with $\Theta_1$ trained above. Initialize parameters in the output layer randomly. $\Theta_2$ is the set of parameters related to the main task \\
\For{iteration in $1, 2, \dots$}{
    Pick a mini-batch from $\mathcal{S}^{SF}$\\
    Compute loss $\mathcal{L}_{main}(\Theta_2)$ according to Eq.(\ref{eq.7.})\\
    Compute gradient: $\nabla \mathcal{L}_{main}(\Theta_2)$\\
    Update model: $\Theta_2=\Theta_2-\varepsilon\nabla \mathcal{L}_{main}(\Theta_2)$
}
\end{algorithm}

\section{Implementation Details}
\subsection{Hyper-parameters}
The settings of hyper-parameters are as follows: (1) Embedding layers are randomly initialized using a normal distribution around 0 with standard deviation $10^{-3}$ for each group. For the other layers, parameters are initialized with a uniform distribution in the range $0.036*(-sqrt(3)/sqrt(dim),sqrt(3)/sqrt(dim))$, where $dim$ is the size of the input. (2) The embedding sizes of keywords and features are set to 50 and 24 respectively. (3) The mini-batch size is set to 256 and AdaGrad \cite{duchi2011adaptive} is used as the optimizer, with learning rate set to 0.03. (4) The layer sizes of the two FC layers are set as 256-256. (5) In the training stage, for both pre-training and alternate training the max numbers of epochs are set to 6 and 15 for auxiliary task and main task respectively to avoid overfitting. For the alternate training strategy shown in Algorithm 1, we empirically set the probability proportion parameter $k=4$ which leads to a good performance for the main task. For alternate training, the training will stop when the max number of epochs of either task is reached. The above parameters are set so as to balance the impact of main/auxiliary task, and also make sure in alternate training the model is sufficiently trained with the main task.

\subsection{Hardware and Software}
The proposed models are implemented on TensorFlowRS (TFRS) provided by Alibaba. TFRS is a distributed deep learning platform based on TensorFlow 1.7 used internally in Alibaba. In experiments, the trainable parameters are distributed on 200 workers (20 CPU cores for each worker) and updated asynchronously. 

\section{Dataset Processing}
\noindent \textbf{SF Dataset:} For the main task, we construct the SF dataset from TaoBao App. As shown in Fig. \ref{sp-example2} in the main text, these SFs are displayed for query users. Because the SF dataset contains only one-class implicit feedbacks (i.e., only users' click actions as positive feedbacks), negative sampling is required for obtaining a certain number of un-clicked SFs as negative feedbacks. We randomly sample from the un-clicked SFs to generate negative feedbacks. The ratio of positive feedbacks to negative feedbacks is kept to 1:2, which is determined by cross validation in preliminary offline experiments. For offline evaluation, the SF dataset is randomly split into training and test sets with the ratio of 9:1.

\vspace{3mm} \noindent \textbf{AD Dataset:} For the auxiliary task, we also construct the AD dataset from TaoBao APP, in a similar fashion with the SF dataset. Specifically, we treat the clicked ads as positive feedbacks and sample un-clicked ads before the last clicked ad in a search session as negative feedbacks. In order to reduce the impact of other factors of ads, we collect only the click data for ads equipped with SP exhibition (generated by the basic model). The ratio of positive feedbacks to negative feedbacks for the AD dataset is kept to 1:6.

For both datasets, we also collect rich information about users and queries for feature. These features represent user preference in many ways.


\section{Details of Features}

\noindent \textbf{The Basic Features:} The basic features for users are user preferred words. For each user, we extract his/her click history within the recent 1 month for representation construction. Specifically, we collect keywords from the titles of those products that a user have clicked and use (at most) top 10 frequent keywords to represent the long-term interest of that user; we also extract (at most) top 10 frequent keywords from a user's product click data within the recent one week to represent his/her short-term interest.The basic features for query, ad and SF are query keywords, ad titles and SF keywords, respectively.

\vspace{3mm} \noindent \textbf{The Additional Features:} There are four groups of additional features, 3 groups for users and 1 group for queries (Table \ref{Tab.feature} in the main text). Here we describe them in details.
\begin{itemize}
    \item \textbf{User profile information:} this group contains users' demographic features, i.e. gender, age, occupation and home address. \textit{Gender} is a binary feature. \textit{Age} is a numerical feature, so we discretize it into 10 discrete states, i.e. $[1,10]$, $[11,20]$, $\dots$, $[91,100]$. \textit{Occupation} is a categorical feature with about 140 occupations in Taobao. \textit{City} and \textit{Province} are also categorical features containing cities and provinces in China.
    
    \item \textbf{User general preference:} this group measures a user's preference regarding general aspects. \textit{Preference for categories} and \textit{Preference for brands} are BoW features. For the former one, we collect ads that a user have clicked, collected or bought, and count the number of ads for each category. Finally, we use (at most) top 10 frequent categories to represent that user's preference for categories. There are totally about 40,000 categories in the dataset. The feature generation process for the latter one is the same. There are about 20,000 brands in the dataset. Finally, \textit{Preference for discount} is a binary feature indicating whether the user likes discounts.
    
    \item \textbf{User consumption/activity level:} \textit{Purchase level} and \textit{Vip level} are both discrete feature with 7 states used in Taobao internally. Purchase level is estimated from and positively correlated with a user's recent and historical consumptions. VIP level considers not only consumption level, but also activity level (e.g., frequency of writing reviews) of a user. \textit{High consumption visitors} and \textit{top class visitors} are binary variables indicating top users. High consumption means a user achieves a very high consumption level; ``top class'' is a service in Taobao that can be bought by high consumption users.
    
    \item \textbf{Query category:} The \textit{Category} feature is a BoW feature. In the search engine of Taobao, a query is matched to a specific category (in case an ambiguity exists, we take all the matched categories) in a hierarchy. We employ the matched category and its ancestors as a query's category information. In the dataset, there are about 400,000 categories.

\end{itemize}

\end{document}